\documentclass[11pt, twocolumn]{aastex631}
\usepackage{hyperref}
\newcommand{\kms}{${\rm km~s^{-1}}$}
\newcommand{\Moyr}{$M_{\odot}~{\rm yr^{-1}}$}
\newcommand{\Ha}{${\rm H\alpha}$}
\newcommand{\Hb}{${\rm H\beta}$}

\begin{document}
	
\title{\textbf{\large{Enhanced Star Formation Activity of Extreme Jellyfish Galaxies in Massive Clusters and the Role of Ram Pressure Stripping 
}}}

\author[0000-0003-3301-759X]{Jeong Hwan Lee}
\affil{Astronomy Program, Department of Physics and Astronomy, SNUARC, Seoul National University, 1 Gwanak-ro, Gwanak-gu, Seoul 08826, Republic of Korea}
\author[0000-0003-2713-6744]{Myung Gyoon Lee}
\affil{Astronomy Program, Department of Physics and Astronomy, SNUARC, Seoul National University, 1 Gwanak-ro, Gwanak-gu, Seoul 08826, Republic of Korea}
\author[0000-0002-3706-9955]{Jae Yeon Mun}
\affil{Research School of Astronomy and Astrophysics, Australian National University, Canberra, ACT 2611, Australia}
\author{Brian S. Cho}
\affil{Astronomy Program, Department of Physics and Astronomy, SNUARC, Seoul National University, 1 Gwanak-ro, Gwanak-gu, Seoul 08826, Republic of Korea}
\author[0000-0003-3734-1995]{Jisu Kang}
\affil{Astronomy Program, Department of Physics and Astronomy, SNUARC, Seoul National University, 1 Gwanak-ro, Gwanak-gu, Seoul 08826, Republic of Korea}

\correspondingauthor{Myung Gyoon Lee}
\email{joungh93@snu.ac.kr, mglee@astro.snu.ac.kr}
\keywords{Galaxy environments (2029) --- Galaxy clusters (584) --- Ram pressure stripped tails (2126) --- Intracluster medium (858) --- Starburst galaxies (1570) --- Galaxy spectroscopy (2171)}

\begin{abstract}

Jellyfish galaxies are an excellent tool to investigate the short-term effects of ram pressure stripping (RPS) on star formation in cluster environments.
It has been thought that the star formation activity of jellyfish galaxies may depend on the host cluster properties, but previous studies have not yet found a clear correlation.
In this study, we estimate the \Ha-based star formation rates (SFRs) of five jellyfish galaxies in massive clusters ($\sigma_{v, {\rm cl}}\gtrsim1000~{\rm km~s^{-1}}$) at $z\sim0.3-0.4$ using Gemini GMOS/IFU observations to explore the relationship.
Combining our results with those in the literature, we find that the star formation activity of jellyfish galaxies shows a positive correlation with their host cluster velocity dispersion as a proxy of cluster mass and dynamical states.
We divide the jellyfish galaxy sample into two groups with strong and weak RPS signatures using a morphological class.
In the phase-space diagram, the jellyfish galaxies with strong RPS features show a higher SFR and a stronger central concentration than those with weak RPS features.
We estimate their SFR excess relative to the star formation main sequence (starburstiness; $R_{\rm SB}={\rm SFR/SFR_{MS}}(z)$) and the density of the surrounding intracluster medium (ICM) using scaling relations with the cluster velocity dispersion.
As a result, the starburstiness of jellyfish galaxies with strong RPS signatures clearly exhibits positive correlations with cluster velocity dispersion, ICM density, and strength of ram pressure.
This shows that the relation between RPS and star formation activity of jellyfish galaxies depends on the host cluster properties and strength of ram pressure.

\end{abstract}


\section{Introduction}

A majority of gas-rich galaxies in galaxy clusters undergo ram-pressure stripping \citep[RPS;][]{gun72}, which is the hydrodynamic interaction of the gas content in a galaxy with the intracluster medium (ICM).
RPS effectively removes gas from cluster galaxies, but it can temporarily induce 
star formation activity in the galaxies.
The stripped gas from the galaxies can be compressed by ram pressure, leading to its collapse and to the formation of new stars in the wake of RPS.
This occurs within a few hundred Myr, as reproduced by simulations \citep{bek03, kro08}.
This process can generate galaxies with jellyfish-like morphologies, showing disturbed tails and extraplanar star-forming knots \citep{ebe14, pog16}.
These jellyfish galaxies are important targets exhibiting a snapshot of starburst galaxies undergoing RPS.

Recent observations have revealed that jellyfish galaxies show systematically enhanced star formation activity compared to normal star-forming galaxies.
Using the sample from the GAs Stripping Phenomena (GASP) survey ($z=0.04-0.07$), \citet{vul18} presented that the jellyfish galaxies show higher star formation rates (SFRs) in their disks by 0.2 dex compared to the control sample without RPS.
In addition, observational results for jellyfish galaxies in the A901/2 \citep{rom19}, A1758N \citep{ebe19}, Coma \citep{rob20}, the clusters from DAFT/FADA and CLASH surveys \citep{dur21}, and A1367 \citep{ped22} have been in agreement with their star formation enhancements.

The star formation enhancement of jellyfish galaxies is expected to be closely related to 
the host cluster properties such as cluster mass, cluster dynamics, or ICM density.
Previous simulations predicted that the star formation activity of gas-rich galaxies could be strongly triggered in environments with high ICM pressure exerted by cluster merger or shocks \citep{kap09, bek10, roe14}.

However, there has been no observational consensus of any explicit correlation between the RPS-induced SFRs and the host cluster properties.
For the GASP sample, \citet{gul20} found no dominant link between tail SFRs and cluster velocity dispersion, suggesting that their stellar mass, position, and velocity also play a role on the SFRs.
This might be because the host clusters of the GASP jellyfish galaxies on average have low cluster velocity dispersion ($\langle\sigma_{v,{\rm cl}}\rangle\sim700$ \kms) and low X-ray luminosity (${\log L_{X}}<44.5~{\rm erg~s^{-1}}$), 
implying that most GASP jellyfish galaxies except for a few extreme ones \citep[like JO201 and JW100;][]{pog19} are likely to experience weak or mild RPS effects with low ICM density.
On the other hand, extreme jellyfish galaxies found in massive merging clusters \citep{owe12, ebe19} would be good examples of vigorous star formation triggered in high ram pressure environments, but quantitative studies of these targets in massive clusters are still lacking.

In this Letter, we address the relation of the SFRs of jellyfish galaxies with host cluster velocity dispersion, ICM density, and strength of ram pressure.
Cluster velocity dispersion is a good tracer of cluster mass and dynamics \citep{mun13}, and it is also known to have a close correlation with 
the X-ray luminosity and the ICM density of the cluster \citep{zha11, gul20}.
We estimate the SFRs of five extreme jellyfish galaxies in the MAssive Cluster Survey (MACS) clusters and Abell 2744 ($\sigma_{v,{\rm cl}}\gtrsim1000~{\rm km~s^{-1}}$) based on Gemini GMOS/IFU observations.
We also combine the \Ha-based SFR values of the known jellyfish samples in the literature in addition to those of our sample, to reveal the relation between SFRs and host cluster properties of the jellyfish galaxies.

This paper is structured as follows.
In {\color{blue} {\bf Section \ref{sec:host}}}, we describe the properties of the host clusters of jellyfish galaxies.
In {\color{blue} {\bf Section \ref{sec:data}}}, we explain the GMOS/IFU data and the methods for analysis.
In {\color{blue} {\bf Section \ref{sec:sfr}}}, we show the SFRs of jellyfish galaxies in relation to stellar mass, cluster velocity dispersion, and phase-space diagrams.
In {\color{blue} {\bf Section \ref{sec:rps}}}, we address the relation of the star formation activity of jellyfish galaxies with the host cluster properties and the degree of RPS.
Throughout this paper, we use the cosmological parameters with $H_{0}=70~{\rm km~s^{-1}~Mpc^{-1}}$, $\Omega_{M}=0.3$, and $\Omega_{\Lambda}=0.7$.

\section{Host Cluster Properties}
\label{sec:host}

\begin{figure}
	\centering
	\includegraphics[width=0.5\textwidth]{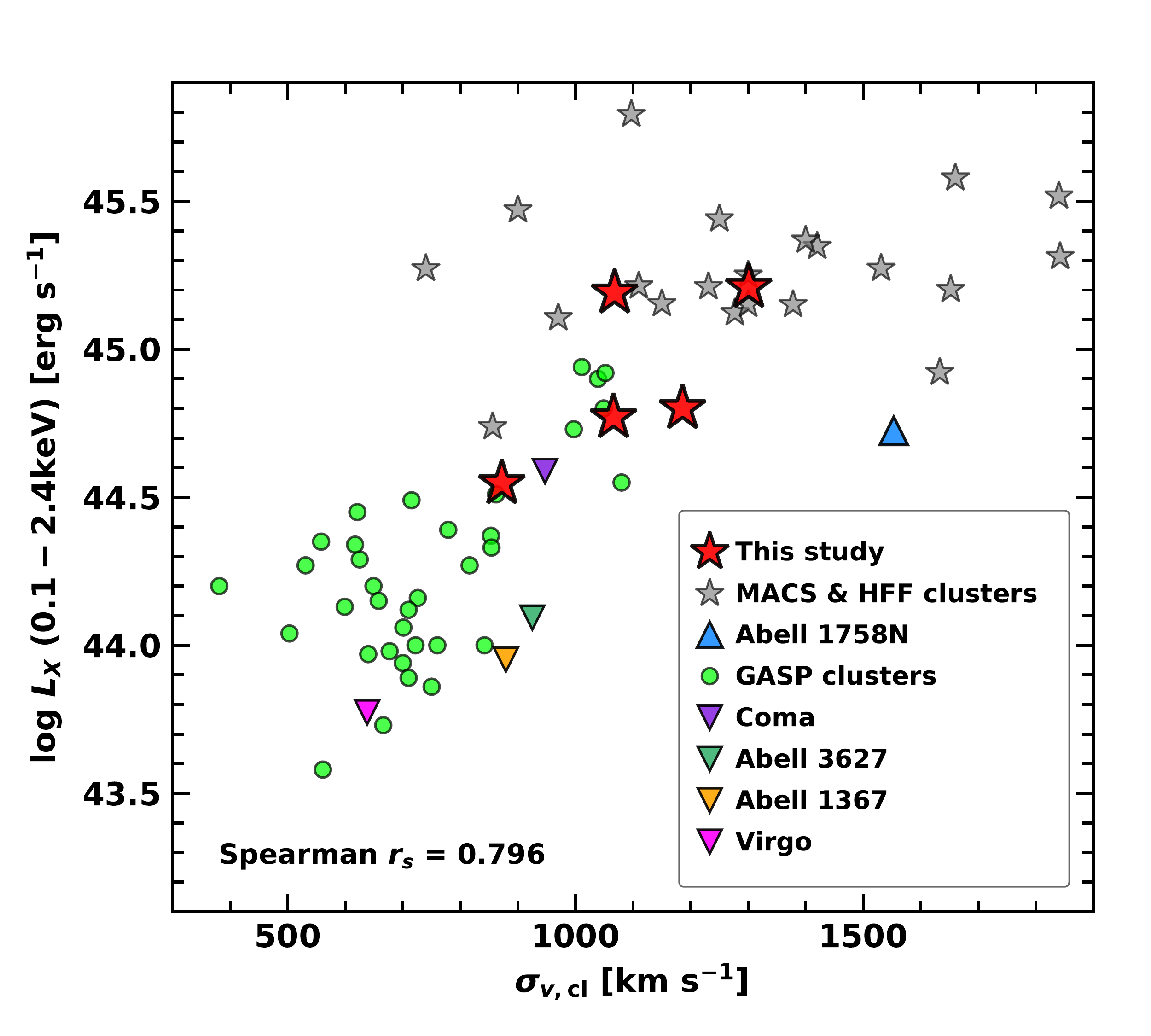}
	\caption{
	Distribution of the X-ray luminosity ($L_{X}$) of the host clusters of jellyfish galaxies as a function of the cluster velocity dispersion ($\sigma_{v,{\rm cl}}$).
	Green circles show the data of clusters observed by the GASP survey.
	Upside-down triangle symbols show several well-known clusters: the Coma cluster (purple), Abell 3627 (green), Abell 1367 (yellow), and the Virgo cluster (magenta).
	Blue triangle shows the data of Abell 1758N \citep{ebe19}.
    Gray star symbols show cluster samples from the MACS and HFF survey \citep{ebe07, lot17, ric21}.
	Red star symbols show the data of the 5 clusters (MACSJ0916.1$-$0023, MACSJ1752.0$+$4440, Abell 2744, MACSJ1258.0$+$4702, and MACSJ1720.2$+$3536) in this study.
	\label{fig:Xray}}
\end{figure}

{\color{blue} {\bf Figure \ref{fig:Xray}}} shows the relation for the host clusters of jellyfish galaxies between the cluster velocity dispersion ($\sigma_{v,{\rm cl}}$) and the X-ray luminosity ($L_{X}$) observed in the energy range of $0.1-2.4~{\rm keV}$.
The X-ray data of the clusters were obtained from the ROSAT All-Sky Survey \citep{boe96, ebe98, vog99}.
We plot the data of the GASP clusters \citep{pog16, gul20}, 4 nearby clusters \citep[Coma, Abell 3627, Abell 1367, and Virgo;][and references therein]{bos21}, Abell 1758N \citep{ebe19}, and the MACS and HFF clusters \citep{ebe07, lot17, ric21}, including the host clusters of five extreme jellyfish galaxies (red star symbols).
The MACS and HFF clusters show much higher velocity dispersion and X-ray luminosity than the nearby clusters.
In comparison with the GASP clusters ($\langle\sigma_{v, {\rm cl}}\rangle=731~{\rm km~s^{-1}}$),
the MACS and HFF clusters have a much higher mean velocity dispersion ($\langle\sigma_{v, {\rm cl}}\rangle=1296~{\rm km~s^{-1}}$).
In addition, most of the GASP clusters show lower X-ray luminosity than $\log L_{X}=44.5~{\rm erg~s^{-1}}$, but all the clusters from the MACS and HFF show $\log L_{X}>44.5~{\rm erg~s^{-1}}$.
This indicates that massive clusters like the MACS and HFF clusters have a much denser ICM than the nearby low-mass clusters.
In addition, these massive clusters tend to be dynamically unstable with cluster collisions or major mergers, exerting shocks and increasing 
ram pressure to their member galaxies \citep{man12, owe12}.
Thus, the 
five extreme jellyfish galaxies in the MACS clusters and Abell 2744 
are expected to suffer from a much stronger degree of RPS compared to the local jellyfish galaxies such as the GASP sample.
This can be also supported by the results from \citet{mor22}, which showed that jellyfish galaxies in the central region of the two HFF clusters (Abell 2744 and Abell 370) are undergoing strong RPS.

\section{Data and Methods}
\label{sec:data}

\subsection{Observations and Data Reduction}
We observed 
five jellyfish galaxies (MACSJ0916-JFG1, MACSJ1752-JFG2, A2744-F0083, MACSJ1258-JFG1, and MACSJ1720-JFG1) 
during four GMOS/IFU observation programs from March 2019 to June 2021.
These jellyfish galaxies were first reported in \citet{owe12} and \citet{mcp16}.
We used the 2-slit mode with the field-of-view (FOV) of $5''\times7''$ and the gratings of R400 (A2744-F0083) and R150 (the others).
The science exposure times ranged from 1.2 hr to 4.2 hr.
All the obtained GMOS/IFU data covered at least the \Ha+[\ion{N}{2}] regions.
These GMOS/IFU data were reduced with the \texttt{PyRAF/Gemini} package and combined with a pixel scale of $0\farcs1~{\rm pixel^{-1}}$.
The detailed reduction process will be given in Lee et al. (2022, in preparation).

\subsection{Emission Line Analysis and SFRs}

SFRs were derived from \Ha~luminosity corrected for stellar absorption and dust extinction.
We carried out Gaussian smoothing of GMOS/IFU spectra with masking emission lines and subtracted the smoothed continuum from the spectra.
We then adopted the \citet{car89} dust extinction laws and the \citet{cha03} initial stellar mass function (IMF), as used in the GASP studies.
Since this study collects and compares the \Ha-based SFR values of jellyfish galaxies in the A901/2 \citep[][RO19 hereafter]{rom19} and A1758N \citep[][EK19 hereafter]{ebe19}, we also converted their SFR values to those 
for \citet{cha03} IMF for consistency.

The spaxels with ${\rm S/N~(H\alpha)}<3$ or AGN/LINER emission in the BPT diagrams ([\ion{O}{3}]$\lambda5007$/\Hb~vs. [\ion{N}{2}]$\lambda6584$/\Ha) are excluded for computing SFRs.
If the \Hb+[\ion{O}{3}] region is out of the wavelength coverage or has a lower S/N than 3 in the spectra, we only regarded the spaxels with log([\ion{N}{2}]$\lambda6584/{\rm H\alpha})<-0.4$) as star-forming ones \citep{med18}.
Using these criteria, the spaxels in the central region ($R\lesssim1\arcsec$) of two galaxies (A2744-F0083 and MACSJ1258-JFG1) are classified as the AGN/LINER region.
Lee et al. (2022, in preparation) will present the detailed methods for emission line analysis and give the computed values of SFRs.

We also divided each jellyfish galaxy into the disk and tail regions, 
using the same definition as in the GASP study \citep{pog19} to calculate the total SFR, the tail SFR, and the tail SFR fraction ($f_{\rm SFR}={\rm SFR(tail)/SFR(total)}$).
Unlike the MUSE IFU data used in the GASP studies, our GMOS/IFU spectra have too low S/N to perform the spectral continuum fitting.
Instead, we estimated stellar masses of the jellyfish galaxies from their NIR fluxes within the GMOS/IFU FOV.

\subsection{Strength of Ram Pressure}
\label{sec:ram}
The ram pressure on a galaxy can be computed with $P_{\rm ram}=\rho_{\rm ICM}\times\Delta v_{\rm 3D}^2$, where $\rho_{\rm ICM}$ is the ICM density and $\Delta v_{\rm 3D}^2$ is the 3D relative velocity of the galaxy with respect to the surrounding ICM \citep{gun72}.
For the ICM density, we assumed the static ICM $\beta$-model:
\begin{equation}
    \rho_{\rm ICM}(r_{\rm cl,3D})=\rho_{0}\times\left[ 1+ \left( \frac{r_{\rm cl,3D}}{R_{c}} \right)^{2} \right]^{-3\beta/2},
\end{equation}
where $\rho_{0}$ is the ICM density at the cluster center, $r_{\rm cl,3D}$ is the 3D clustercentric distance, and $R_{c}$ is the core radius of the host cluster.
We assumed $\beta=0.5$ and adopted Equation 16 in \citet{gul20} to derive ICM density from cluster velocity dispersion.
We converted the projected clustercentric distance ($R_{\rm cl}$) and the line-of-sight velocity ($\Delta v_{\rm los}$) to the 3D parameters ($r_{\rm cl,3D}$ and $\Delta v_{\rm 3D}$) by multiplying a factor of $\pi/2$ and $\sqrt{3}$, respectively \citep{jaf18}.

\begin{figure*}
	\centering
	\includegraphics[width=0.7\textwidth]{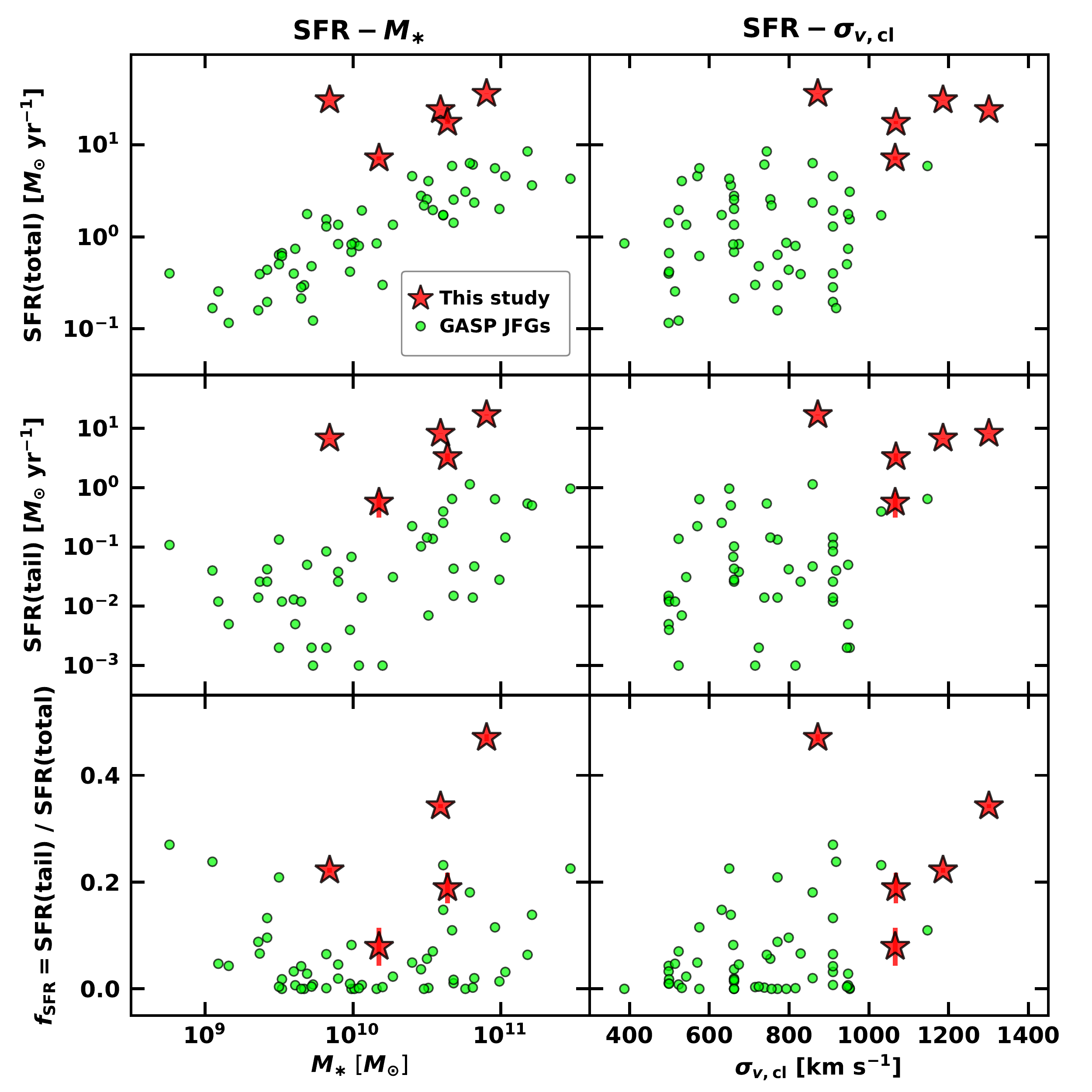}
	\caption{
	Total SFR (upper), tail SFR (middle), and the tail SFR fraction ($f_{\rm SFR}$; lower) as a function of stellar mass (left) and cluster velocity dispersion (right).
	We plot our data (red star symbols) and 54 jellyfish galaxies observed by the GASP survey (green circles) for comparison.
	\label{fig:G20SFR}}
\end{figure*}

There are several caveats of this method.
First, the static ICM $\beta$-model might be difficult to be applied to clusters undergoing collisions or mergers.
For example, merging clusters such as MACSJ1752.0+4440 and Abell 2744 exhibit a disturbed X-ray 
morphology \citep{owe11, fin21}, implying that the ICM distribution is not homogeneous.
Second, the scaling relations in \citet{gul20} might have non-negligible scatter.
These relations were derived from a simple linear interpolation of two model clusters (a low-mass cluster and a high-mass cluster) from Table 1 in \citet{jaf18}.
Thus, the relations could be oversimplified for estimating the ICM density in clusters with a wide range of virial masses.
Third, the projection effect could lead to scatter.
Despite these limitations, we roughly computed the strength of ram pressure of jellyfish galaxies to investigate the relation between the star formation activity and the degree of RPS in {\color{blue} {\bf Section \ref{sec:rps}}}.

\section{Star Formation Activity of the Jellyfish Galaxies}
\label{sec:sfr}

\subsection{Comparison of SFRs with the GASP Sample}

In the left panels of {\color{blue} {\bf Figure \ref{fig:G20SFR}}}, we plot the total SFRs (upper), tail SFRs (middle), and $f_{\rm SFR}$ (lower) of our GMOS/IFU sample and the GASP sample as a function of stellar mass.
The stellar mass range 
of our targets in this study 
is $\log M_{\ast}/M_{\odot}=9.8-10.9$, which is comparable to that of the massive GASP jellyfish galaxies.
Total SFRs of the GASP jellyfish galaxies are clearly proportional to stellar mass.
Our targets show a similar trend, but the total SFRs are by a factor of 10 higher than those of the GASP sample in a similar stellar mass range.
The five jellyfish galaxies show a median SFR of 23.8 \Moyr~in total, whereas the GASP sample shows 1.1 \Moyr.
Tail SFRs of the GASP jellyfish galaxies increase as the stellar mass increases in the range of $M_{\ast}>10^{10}~M_{\odot}$.
In the low-mass regime ($M_{\ast}<10^{10}~M_{\odot}$), such trend is not clear due to the large scatter.
Our targets show higher tail SFRs (median = 6.8 \Moyr) than the GASP sample (median = 0.03 \Moyr).
The median $f_{\rm SFR}$ of our sample is 22\%, which is also by a factor of 10 higher than the GASP sample with $f_{\rm SFR}=3\%$.
Overall, the star formation activity of our sample is more enhanced than that of the GASP sample in terms of total SFR, tail SFR, and $f_{\rm SFR}$.

\begin{figure*}
	\centering
	\includegraphics[width=0.76\textwidth]{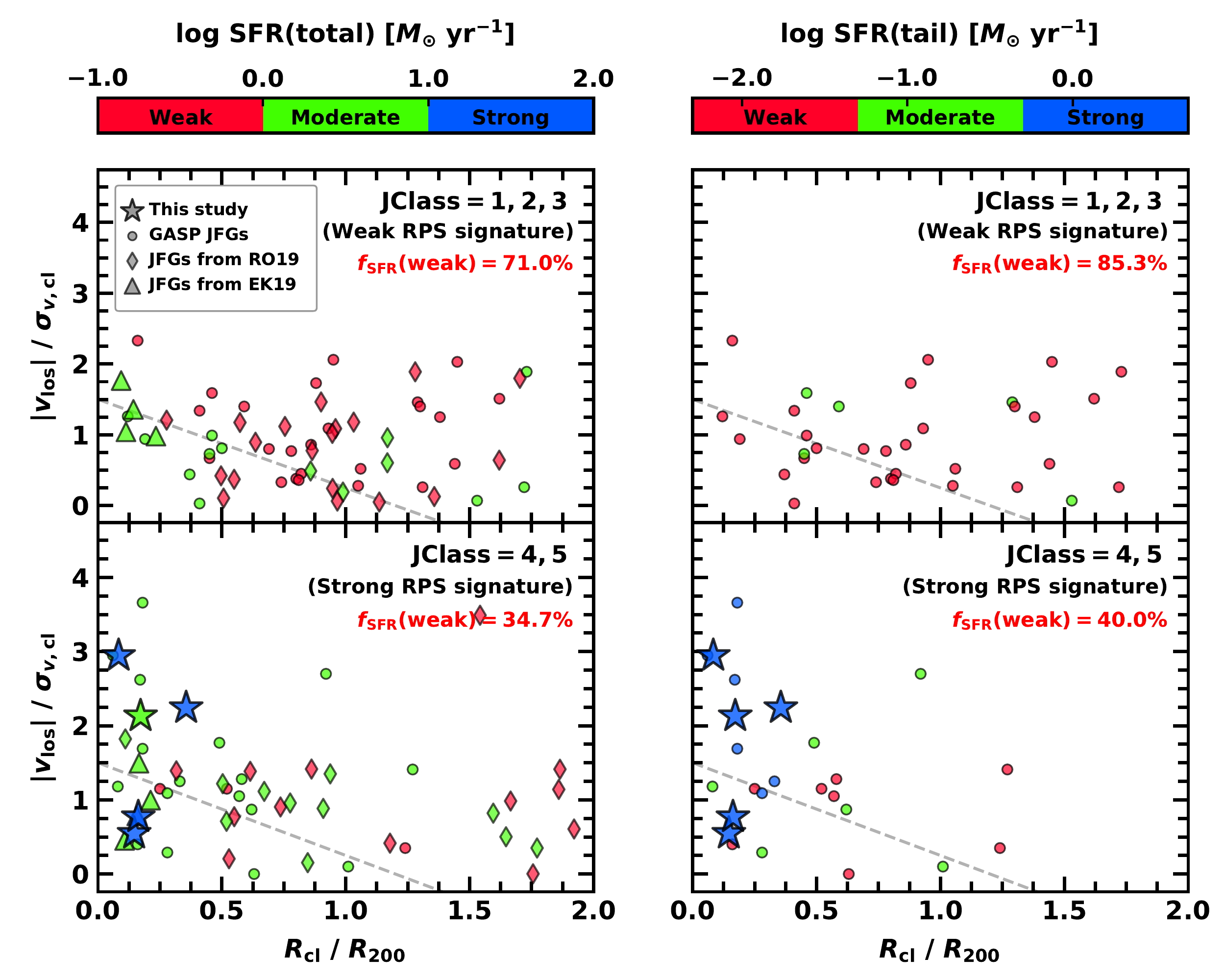}
	\caption{
	Projected phase-space diagrams of our sample (star symbols), the GASP jellyfish galaxies (circles), the A901/2 sample (RO19; diamonds), and the A1758N sample (EK19; triangles).
	We normalize clustercentric distance ($R_{\rm cl}$) and absolute relative velocity ($|v_{\rm los}|$) with cluster virial radius ($R_{200}$) and velocity dispersion ($\sigma_{v, {\rm cl}}$), respectively.
	All the data are color-coded by total SFR (left) and tail SFR (right).
	The color bars on the top denote the logarithmic scale of each SFR, showing the three categories of star formation activity: `weak', `moderate', and `strong'.
	Gray dashed lines represent a boundary of virialized region and recent infall region \citep{jaf15}.
	We divide the whole sample into two categories by JClass from the GASP studies \citep{pog16, jaf18, gul20}: weak RPS signature (JClass = 1, 2, 3; upper) and strong RPS signature (JClass = 4, 5; lower) in the jellyfish galaxies.
	\label{fig:PSDL}}
\end{figure*}

In the right panels, we plot the total SFR, tail SFR, and $f_{\rm SFR}$ versus the cluster velocity dispersion.
The figures show that there is no significant correlation between SFRs (or $f_{\rm SFR}$) and the host cluster velocity dispersion when only the GASP sample is taken into account, as mentioned in \citet{gul20}.
The jellyfish galaxies in this study help us probe higher values of cluster velocity dispersion.
The host clusters of our sample have a median velocity dispersion of $\sigma_{v,{\rm cl}}=1068~{\rm km~s^{-1}}$, which is much higher than that of the GASP clusters (median $\sigma_{v,{\rm cl}}=731~{\rm km~s^{-1}}$).
Combining our data and the GASP sample, we find that the SFRs and $f_{\rm SFR}$ of jellyfish galaxies tend to increase as the cluster velocity dispersion increases.
This implies there may be a positive correlation between the star formation activity and the cluster velocity dispersion in spite of large scatters.
We discuss this correlation further in {\color{blue} {\bf Section \ref{sec:rps}}}.

\subsection{Phase-space Analysis with Jellyfish Morphology}
\label{sec:psd}

\begin{figure*}
	\centering
	\includegraphics[width=0.75\textwidth]{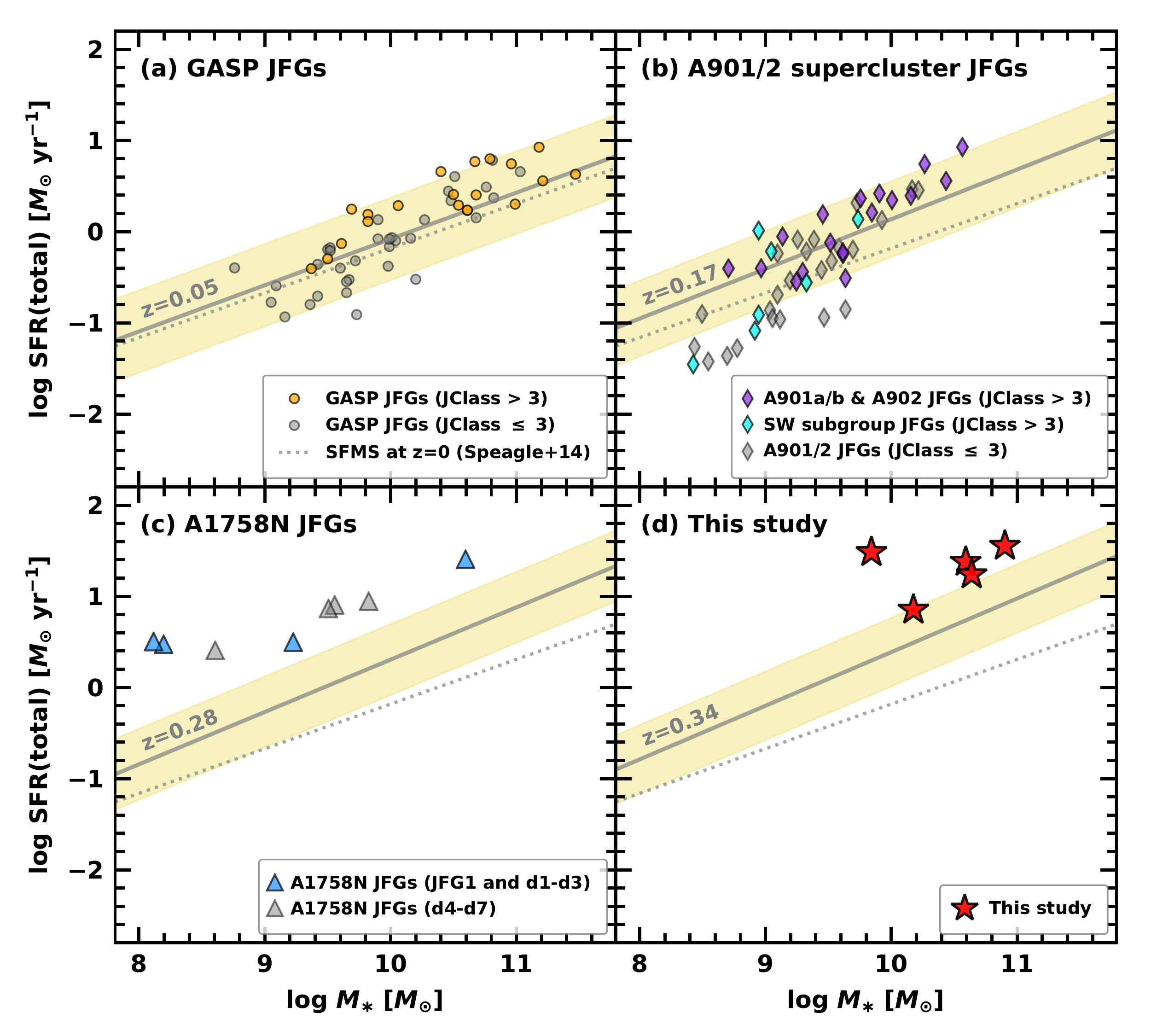}
	\caption{
	The SFR-$M_{\ast}$ diagrams of jellyfish galaxies from the GASP survey (upper left), the A901/2 supercluster (RO19; upper right), A1758N (EK19; lower left), and our GMOS/IFU study (lower right) compared with the star formation main sequence (SFMS) at the median redshift of each sample.
    In the left panel, we mark jellyfish galaxies with strong RPS signatures as colored symbols and those with weak RPS signatures as gray symbols.
    Solid lines and shaded regions show the linear-fit lines of the SFMS and their uncertainty suggested by \citet{spe14}.
    Gray dashed lines denote the linear-fit line of the SFMS at $z=0$.
	\label{fig:SFMS}}
\end{figure*}

In {\color{blue} {\bf Figure \ref{fig:PSDL}}}, we illustrate the projected phase-space diagrams of our targets in addition to samples from the GASP survey \citep{gul20}, A901/2 supercluster (RO19),  
and A1758N (EK19).
We color-code all the jellyfish galaxies with the total SFRs (left panels) and tail SFRs (right panels).
Here we categorize the jellyfish galaxies with the visual classification in \citet{pog16}: JClass = 1, 2, 3 (tentative or probable jellyfish candidates) and JClass = 4, 5 (classical jellyfish galaxies).
The jellyfish galaxies with higher JClass show stronger RPS signatures such as bright tails and extraplanar knots in the optical images or \Ha~flux distributions.
For the GASP sample, the JClass values were given in \citet{gul20}.
RO19 also adopted the JClass as a morphological index of the selected jellyfish sample.
EK19 
classified their sample into galaxies with discernible tails (JFG1 and d1 to d3) and ambiguous RPS features (d4 to d7).
Our GMOS/IFU targets were regarded as classical examples of jellyfish galaxies in previous studies \citep{ebe14, mcp16}, so we classified all our targets as ``strong RPS signature''.

The phase-space diagrams show that the jellyfish galaxies with strong RPS signatures show higher SFRs in total and in tails than those with weak RPS signatures.
Furthermore, the GASP and RO19 samples with strong RPS features are more concentrated on the cluster center than those with weak RPS features (p-value = 0.06 for one-sided Kolmogorov-Smirnov test).
This implies that the jellyfish galaxies with stronger RPS signatures show more enhanced star formation activity compared to those with weaker ones.


\subsection{Comparison of SFRs with the SFMS}
\label{sec:sfms}

\begin{figure*}
	\centering
	\includegraphics[width=1.0\textwidth]{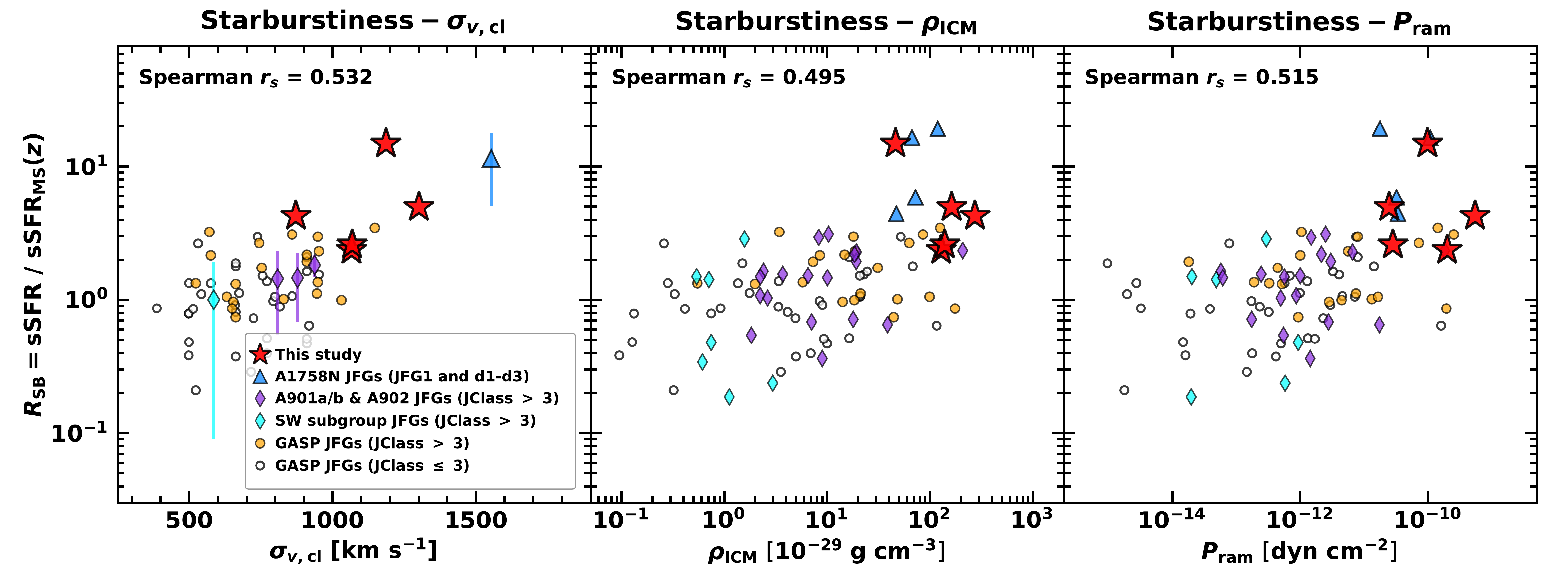}
	\caption{
	Starburstiness ($R_{\rm SB}$) of jellyfish galaxies as a function of cluster velocity dispersion ($\sigma_{v, {\rm cl}}$; left), the ICM density ($\rho_{\rm ICM}$; middle), and the degree of ram pressure ($P_{\rm ram}$; right).
    Error bars in the left panel represent standard deviations of starburstiness of jellyfish galaxies in the same host clusters.
    We plot the data of jellyfish samples with strong RPS signatures as described in {\color{blue} {\bf Figure \ref{fig:PSDL}}}.
    The Spearman's rank correlation coefficients are shown at the top of each panel.
	\label{fig:dSFR}}
\end{figure*}

In {\color{blue} {\bf Figure \ref{fig:SFMS}}}, we plot the integrated SFR-$M_{\ast}$ diagrams of the jellyfish galaxies in comparison with the star formation main sequence (SFMS) at the median redshifts of the jellyfish samples: the GASP galaxies ($z=0.05$; a), the A901/2 sample ($z=0.17$; b), the A1758N jellyfish galaxies ($z=0.28$; c), and our sample ($z=0.34$; d).
We adopted the following SFMS in \citet{spe14} as a function of stellar mass and cosmic time.
\begin{equation}
    \log {\rm SFR}(M_{\ast},~t)=(0.84-0.026\times t)\log M_{\ast}-(6.51-0.11\times t),
\end{equation}
where $t$ is the age of the universe at the redshift of the galaxies in Gyr.
This SFMS model was derived from a compilation of 25 previous studies, most of which studied star-forming galaxies in the field environments.
Note that the SFRs of cluster galaxies could be more suppressed compared to the above SFMS because the SFR-$M_{\ast}$ relation also depends on the environment as shown in the studies of star-forming galaxies at low-$z$ \citep{pac16} and intermediate-$z$ \citep{vul10}.

In the upper panels, we plot the data of the GASP sample (left) and the A901/2 sample (right) whose host systems have on average lower velocity dispersions than 1000 \kms.
The GASP clusters have a mean cluster velocity dispersion of $731~{\rm km~s^{-1}}$, and the 4 subgroups in A901/2 have velocity dispersions of $\sigma_{v,{\rm cl}}=878~{\rm km~s^{-1}}$ for A901a,
$\sigma_{v,{\rm cl}}=937~{\rm km~s^{-1}}$ for A901b, $\sigma_{v,{\rm cl}}=808~{\rm km~s^{-1}}$ for A902, and $\sigma_{v,{\rm cl}}=585~{\rm km~s^{-1}}$ for the southwest (SW) group \citep{wei17}.
For the GASP sample, most jellyfish galaxies with JClass $>$ 3 exhibit higher SFRs than not only those with JClass $\leq$ 3 but also those that lie along the SFMS.
The jellyfish galaxies in the A901/2 supercluster seem to follow a similar trend with the GASP sample.
Furthermore, the jellyfish galaxies with JClass $>$ 3 in more massive subgroups (A901a/b and A902) show higher SFR excess relative to the SFMS than those in the SW group.
These results indicate that the jellyfish sample exhibits more enhanced star formation activity as their RPS features become stronger and their hosts become more massive.

In the lower panels, we plot the data of A1758N sample and our sample in massive clusters ($\sigma_{v,{\rm cl}}\gtrsim1000~{\rm km~s^{-1}}$).
All the jellyfish galaxies of A1758N and ours are located clearly above the SFMS, implying that the jellyfish galaxies in massive clusters tend to 
show more enhanced star formation activity compared to those in the GASP clusters and the A901/2 subgroups.
Thus, the significant enhancement of the star formation activity could be due to the difference in the properties of the host clusters (e.g.~the cluster mass, cluster velocity dispersion, or ICM density) which affects the strength of ram pressure on the jellyfish galaxies.




\section{The Relation between the Star Formation Activity and RPS}
\label{sec:rps}

In this section, we explore how the star formation activity of jellyfish galaxies depends on their host cluster velocity dispersion and the strength of ram pressure.
We estimate the value of starburstiness ($R_{\rm SB}$) of the jellyfish galaxies, defined as a ratio between the specific star formation rate (sSFR) of a galaxy to that of the SFMS at the same redshift, indicative of relative star formation activity with respect to the normal galaxies \citep{elb11}.

{\color{blue} {\bf Figure \ref{fig:dSFR}}} illustrates the starburstiness of the jellyfish galaxies as a function of the host cluster velocity dispersion (left panel), the ICM density (middle panel), and the strength of ram pressure (right panel).
For all the panels, we plot the starburstiness of our sample (star symbols) in addition to the GASP (circles), RO19 (diamonds), and EK19 (triangles) sample with strong RPS signature (JClass $>$ 3)
This selection allows us to compare the star formation activity of jellyfish galaxies with similar morphological classes.

In the left panel, the starburstiness of the GASP and RO19 samples with JClass $>$ 3 does not seem to have a clear correlation with the cluster velocity dispersion.
However, we note that there is a positive correlation between $R_{\rm SB}$ and $\sigma_{v,{\rm cl}}$ by adding the data of our sample and the A1758N sample in massive clusters ($\sigma_{v, {\rm cl}}\gtrsim1000~{\rm km~s^{-1}}$).
The Spearman's rank correlation coefficient ($r_{s}$) is 0.532 (p-value = $3.4\times10^{-5}$), indicating that this correlation is reliable.
In the middle and right panels, this trend similarly appears in the relations of $R_{\rm SB}$ vs. $\rho_{\rm ICM}$ ($r_{s}=0.50$ and p-value = $1.4\times10^{-4}$) and $R_{\rm SB}$ vs. $P_{\rm ram}$ ($r_{s}=0.51$ and p-value = $8.0\times10^{-5}$) because the cluster velocity dispersion is closely related to the ICM density and the strength of ram pressure as described in {\color{blue} {\bf Section \ref{sec:ram}}}.

These results imply that the star formation activity of the jellyfish galaxies with similar morphological classes has positive correlations with the host cluster velocity dispersion and the degree of RPS.
Furthermore, these correlations can be more strengthened considering that the starburstiness of our sample and EK19 sample might be underestimated due to possible suppression of SFRs of the SFMS in the cluster central region \citep{pac16}.
In the previous literature, \citet{gul20} pointed out that the star formation activity of the GASP jellyfish galaxies hardly shows remarkable relations with 
the cluster velocity dispersion.
However, the reliable correlations between star formation activity and RPS could be found in this work thanks to the data of jellyfish galaxies in clusters more massive ($\sigma_{v,{\rm cl}}\gtrsim1000~{\rm km~s^{-1}}$) 
than those in the GASP and RO19 studies.
We interpret that this relation clearly shows the short-term effect of RPS on the star formation activity of jellyfish galaxies in clusters.
Although it is expected that stronger RPS will eventually strip the gas of cluster galaxies, it could trigger the star formation activity more strongly in jellyfish galaxies instead.

\section{Summary}
\label{sec:summary}

In this study, we investigate the relation between the star formation activity of jellyfish galaxies and their host cluster properties.
We use the Gemini GMOS/IFU observations of five extreme jellyfish galaxies in the MACS clusters and Abell 2744 at $z>0.3$ 
for our study.
We computed \Ha-based SFRs and compared them to those from the GASP, RO19, and EK19 samples using the SFR$-M_{\ast}$ and phase-space diagrams.
We summarize our results as follows.

\begin{enumerate}
	\item In the SFR$-M_{\ast}$ and SFR$-\sigma_{v,{\rm cl}}$ diagrams, the total SFRs, tail SFRs, and $f_{\rm SFR}({\rm tail})$ of the five jellyfish galaxies are an order of magnitude higher than those of the GASP jellyfish galaxies.
	Combining our data and the GASP results, the SFRs and $f_{\rm SFR}$ of jellyfish galaxies tend to increase as the stellar mass and cluster velocity dispersion increase.
	\item The projected phase-space diagrams of the combined sample of the GASP survey, RO19, EK19, and ours indicate that jellyfish galaxies with strong RPS signatures (JClass $>$ 3) show more enhanced star formation activity compared to those with weak RPS signatures (JClass $\leq$ 3).
    \item In the SFR-$M_{\ast}$ diagram, our sample and the EK19 sample are located above the SFMS at their median redshifts.
    The SFR excess of our sample and the EK19 sample (massive clusters) is also higher than that of the GASP and RO19 sample (low-mass clusters), implying that the star formation activity of jellyfish galaxies in massive clusters is more enhanced.
    \item Combining all the jellyfish galaxies with strong RPS features, 
    we find that starbustiness correlates positively with the cluster velocity dispersion, ICM density, and strength of ram pressure.
    This implies that jellyfish galaxies show more enhanced star formation activity with increasing host cluster mass and degree of ram pressure.
\end{enumerate}

\begin{acknowledgments}
We would like to thank the anonymous referee for his/her useful comments and suggestions to improve the manuscript.
This study was supported by the National Research Foundation grant funded by the Korean Government (NRF-2019R1A2C2084019).
This work was supported by K-GMT Science Program (PID: GS-2019A-Q-214, GN-2019A-Q-215, GS-2019B-Q-219, and GN-2021A-Q-205) of Korea Astronomy and Space Science Institute (KASI).
Based on observations obtained at the international Gemini Observatory, a program of NSF’s NOIRLab, acquired through the Gemini Observatory Archive at NSF’s NOIRLab and processed using the Gemini IRAF package, which is managed by the Association of Universities for Research in Astronomy (AURA) under a cooperative agreement with the National Science Foundation on behalf of the Gemini Observatory partnership: the National Science Foundation (United States), National Research Council (Canada), Agencia Nacional de Investigaci\'{o}n y Desarrollo (Chile), Ministerio de Ciencia, Tecnolog\'{i}a e Innovaci\'{o}n (Argentina), Minist\'{e}rio da Ci\^{e}ncia, Tecnologia, Inova\c{c}\~{o}es e Comunica\c{c}\~{o}es (Brazil), and Korea Astronomy and Space Science Institute (Republic of Korea).
\end{acknowledgments}

\software{Numpy \citep{har20}, Matplotlib \citep{hun07}, Scipy \citep{vir20}, Astropy \citep{ast13, ast18}, and PyRAF \citep{sts12}}

\clearpage

\end{document}